\documentclass[sn-mathphys,Numbered]{sn-jnl}

\usepackage{subfig}
\usepackage{lipsum}
\usepackage{float}
\usepackage{graphicx}%
\usepackage{multirow}%
\usepackage{amsmath,amssymb,amsfonts}%
\usepackage{amsthm}%
\usepackage{mathrsfs}%
\usepackage[title]{appendix}%
\usepackage{xcolor}%
\usepackage{textcomp}%
\usepackage{manyfoot}%
\usepackage{booktabs}%
\usepackage{algorithm}%
\usepackage{algorithmicx}%
\usepackage{algpseudocode}%
\usepackage{listings}%
\usepackage[numbers]{natbib}
\begin{document}
	
\title[Simulations of Magnetization Reversal in FM/AFM Bilayers With THz Frequency Pulses]{Simulations of Magnetization Reversal in FM/AFM Bilayers With THz Frequency Pulses}

\author*[1]{\fnm{Joel} \sur{Hirst}}\email{joel.hirst@student.shu.ac.uk}
\author[1,2]{\fnm{Sergiu} \sur{Ruta}}
\author[3]{\fnm{Jerome} \sur{Jackson}}
\author[4]{\fnm{Thomas} \sur{Ostler}}

\affil[1]{\orgdiv{Materials \& Engineering Research Institute}, \orgname{Sheffield Hallam University}, \orgaddress{\street{Howard Street}, \city{Sheffield}, \postcode{S1 1WB}, \country{United Kingdom}}}
\affil[2]{\orgdiv{The Department of Engineering and Mathematics}, \orgname{Sheffield Hallam University}, \orgaddress{\street{Howard Street}, \city{Sheffield}, \postcode{S1 1WB}, \country{United Kingdom}}}
\affil[3]{\orgdiv{Scientific Computing Department}, \orgname{ STFC Daresbury Laboratory}, \orgaddress{\city{Warrington}, \postcode{ WA4 4AD}, \country{United Kingdom}}}
\affil[4]{\orgdiv{Department of Physics \& Mathematics}, \orgname{University of Hull}, \orgaddress{\city{Hull}, \postcode{HU6 7RX}, \country{United Kingdom}}}
\abstract{
It is widely known that antiferromagnets (AFMs) display a high frequency response in the terahertz (THz) range, which opens up the possibility for ultrafast control of their magnetization for next generation data storage and processing applications. However, because the magnetization of the different sublattices cancel, their state is notoriously difficult to read. One way to overcome this is to couple AFMs to ferromagnets - whose state is trivially read via magneto-resistance sensors. Here we present conditions, using theoretical modelling, that it is possible to switch the magnetization of an AFM/FM bilayer using THz frequency pulses with moderate field amplitude and short durations, achievable in experiments. Consistent switching is observed in the phase diagrams for an order of magnitude increase in the interface coupling and a tripling in the thickness of the FM layer. We demonstrate a range of reversal paths that arise due to the combination of precession in the materials and the THz-induced fields.  Our analysis demonstrates that the AFM drives the switching and results in a much higher frequency dynamics in the FM due to the exchange coupling at the interface. The switching is shown to be robust over a broad range of temperatures relevant for device applications.
}

\maketitle

\section{Introduction}
\begin{figure*}[!t]
    \centering
        \includegraphics[width=1.0\linewidth]{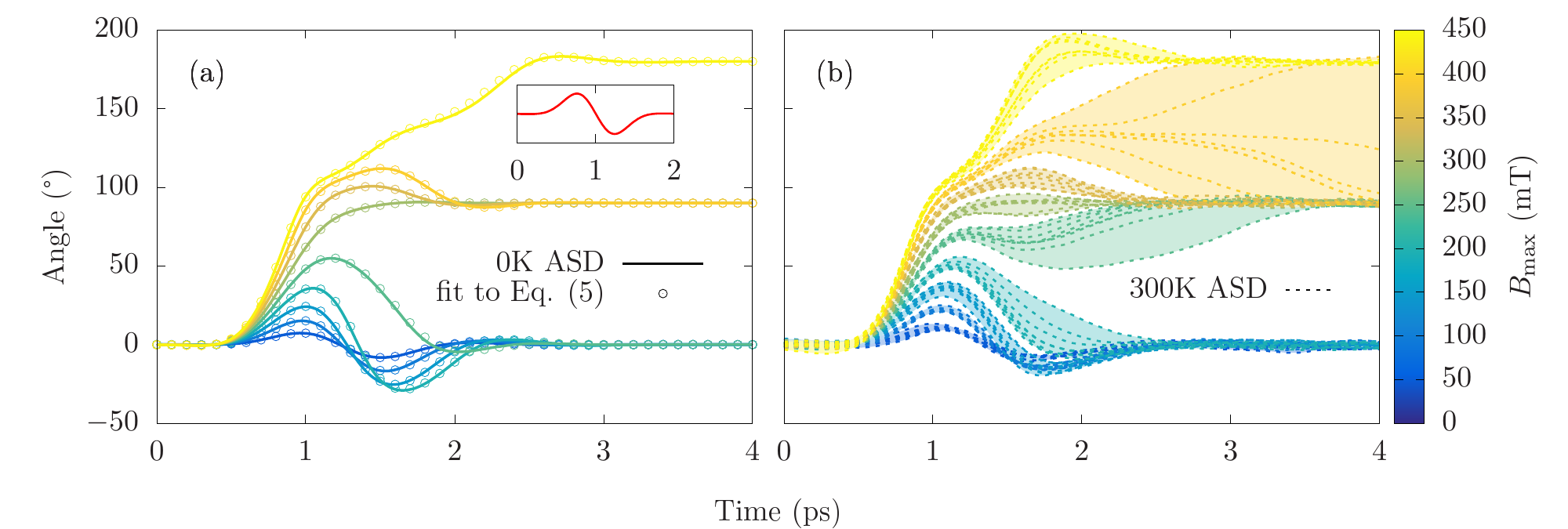}
    \caption{Magnetization reversal in Mn\textsubscript{2}Au for varying maximum field amplitudes, $B_{\mathrm{max}}$, from approximately 50 to 450 mT at (a) $T = 0$ and (b) $T = 300$ Kelvin for a constant pulse width and frequency of $\sigma = 0.3$ ps and $f = 0.67 $ THz. Inset in (a) shows the field profile between 0 and 1 ps or a $\sigma = 0.3$ ps. The dotted lines in (b) represent sublattice dynamics from individual simulations, the shaded area corresponds to the upper and lower bound for each field strength.}
    \label{fig:angle_300K}
\end{figure*}
Antiferromagnets (AFM) are materials with alternating magnetic moments at the atomic level resulting in no net magnetization. Their inherently fast THz magnetization dynamics \cite{kittel1, kittel2} and absence of stray fields potentially means faster magnetization reversal and higher information density which makes them strong candidates for next generation spintronic and magnetic recording media. Due to the antiferromagnetic order, however, controlling them remains a challenge as there is no net Zeeman energy when placed in a magnetic field, which complicates device development, although this is starting to be overcome by taking advantage of spin-orbit effects, e.g. spin-orbit torques.
\\
\\
The use of current-induced N\'eel spin-orbit torques (NSOT) has been shown to be able to induce magnetization dynamics in several AFMs. Two materials, CuMnAs \cite{MACA20121606, Olejnik2017, PhysRevLett.118.057701, Wadley2015} and Mn\textsubscript{2}Au \cite{Bodnar2018, PhysRevApplied.9.064040, PhysRevApplied.16.014037, Bommanaboyena2021}, have been identified with a strong spin-orbit coupling that creates sufficient NSOTs for manipulation of the N\'eel vector. Experiments so far have not taken advantage of the intrinsically fast dynamics of AFMs, with switching via electrical control being limited by the low frequency mode of the antiferromagnet \cite{Wadley2016}. Computationally, It has been shown that square staggered fields can lead to reversal in Mn\textsubscript{2}Au \cite{selzer_mn2au, roy_mn2au, unai_mn2au} with pulse durations of the order of a few picoseconds. While a square field is a good first approximation, experimental realisation of such a profile is challenging. However, advances in the experimental generation of ultrashort THz pulses \cite{thz1, thz2, thz3, thz4} has opened the possibility for excitation and reversal of the N\'eel vector using sub-picosecond single and multi-cycle fields. 
\\
\\
Even though the manipulation of the order parameter can be carried out electrically via NSOTs, the opposing net zero magnetization in AFMs means the readout of the order parameter, otherwise known as the N\'eel vector, still remains a challenge. The opposing sublattices mean that detection of the magnetic order using conventional methods such as the anisotropic magnetoresistance mechanism (AMR) often yields weak signal outputs compared to typical ferromagnetic materials \cite{Wadley2016,PhysRevApplied.14.014004}. However, a recent demonstration of 100\% magnetoresistance in an antiferromagnetic stack of MnPt and Mn$_3$Pt~\cite{Qin2023} which could pave the way for future antiferromagnetic spintronic devices. Another approach of inferring the N\'eel vector is to couple AFMs directly to FM materials and measure the orientation of the AFM through measurements of the magnetization state of the FM. Recently, it has been shown Mn\textsubscript{2}Au can be coupled ferromagnetically at the interface to Permalloy with one-to-one imprinting of the AFM order on the FM domains \cite{Bommanaboyena2021}, though little is known of the dynamics of switching and whether one can simultaneously take advantage of the higher frequency dynamics of the AFM whilst being coupled to the sluggish FM.
\\
\\
In this work, we conduct atomistic spin dynamics (ASD) simulations of pure Mn\textsubscript{2}Au as well as an FM/AFM bilayer consisting of a Mn\textsubscript{2}Au coupled to ferromagnetic Permalloy (Py). Sec. \ref{sec:methods} introduces the atomistic model, the applied magnetic field profile and material properties used for Mn\textsubscript{2}Au. ASD simulations of reversal in Mn\textsubscript{2}Au as a result of a THz frequency field at 0 K and 300 K are then presented in Sec. \ref{sec:mn2au} where we show that, due to the tetragonal anisotropy in the plane, the magnetization can be reoriented along the $[110]$, $[\bar{1}10]$, $[1\bar{1}0]$ and $[\bar{1}\bar{1}0]$ directions.  We show that staggered linearly polarized fields of the order of $\sim 1$ Tesla are required for deterministic reversal. In Sec. \ref{sec:mn2au_py} we determine phase diagrams for magnetization reversal in Mn\textsubscript{2}Au/Py bilayers with varying interface exchange and Py film thicknesses. We compare cases of excitation of both the Mn\textsubscript{2}Au and Py sublattices to excitation of Mn\textsubscript{2}Au alone to probe the relative roles of each layer in the switching. Finally, we simulate switching at 300 K and 600 K to determine whether the temperature scaling of the anisotropy affects the field strengths and pulse durations required for switching.
\section{Methods}
\label{sec:methods}
The spin dynamics of the magnetic moments is described by the well known stochastic Landau-Lifshitz-Gilbert (LLG) equation:
\begin{equation}
    \frac{\partial \boldsymbol{S}_{i}}{\partial t}=-\frac{\gamma_{i}}{\left(1+\lambda_{i}^{2}\right) \mu_{i}}\left(\boldsymbol{S}_{i} \times \boldsymbol{H}_{i}+\lambda_{i} \boldsymbol{S}_{i} \times \boldsymbol{S}_{i} \times \boldsymbol{H}_{i}\right)
    \label{eq:llg}
\end{equation}
where $\mathbf{S}_i$ is a normalised unit vector of the spin at site $i$, $\lambda_i$ is the effective damping parameter, $\gamma = 1.76 \times 10^{-11}$ is the gyromagnetic ratio, $\mu_i$ is the atomic magnetic moment and $\mathbf{H}_i$ is the effective field acting on the spin at site $i$. The effective field is given by the equation:
\begin{equation}
    \mathbf{H}_i=\boldsymbol{\zeta}_i(t)-\frac{\partial \mathcal{H}}{\partial \mathbf{S}_i} + B(t)
\end{equation}
where $\boldsymbol{\zeta}_i(t)$ describes the coupling to the thermal bath and $B(t)$ is the applied external field. For atomistic simulations of pure Mn\textsubscript{2}Au, the following Heisenberg Hamiltonian is used:
\begin{equation}
\mathcal{H}_{\textrm{Mn}}= \sum_{i \neq j} J_{i j}^{\textrm{Mn}} \boldsymbol{S}_i \cdot \boldsymbol{S}_j-\sum_i d_z S_{i, z}^2 -\sum_i d_{z z} S_{i, z}^4-\sum_i d_{x y} S_{i, x}^2 S_{i, y}^2
\label{eq:hamil_Mn}
\end{equation}
with anisotropy constants taken from Ref. \cite{selzer_mn2au} $d_z=-1.19$ meV, $d_{z z}=-0.015$ meV and $d_{x y}=0.04$ meV following the original calculations in Ref. \cite{shick_mn2au}. We use a magnetic moment of $\mu_{\textrm{Mn}} = 3.87 \mu_B$. The exchange interactions are taken from Ref. \cite{hirst_mn2au}.  We use a value of $\lambda = 0.01$ in Mn\textsubscript{2}Au as seen in previous atomistic modelling \cite{selzer_mn2au, hirst_mn2au, roy_mn2au}. The Mn sublattices are initialised along the [1,-1,0] and [-1,1,0] directions respectively. The LLG equation is integrated using the Heun method \cite{Nowak2007}.
\\
\\
The staggered field, $B(t)$, is applied along the perpendicular direction in the easy-plane to maximise the torque. Such staggered fields can be generated in metallic antiferromagnets such as Mn\textsubscript{2}Au as a result of electrical currents \cite{Salemi2019,PhysRevApplied.9.064040,PhysRevLett.113.157201} . The field is modelled as a Gaussian envelope with fixed frequency of 0.67 THz, which is close to the in-plane resonant resonant mode. The field equation is given by:
\begin{equation}
B(t)= H \exp\left({-\frac{\left(t-t_0\right)^2}{2 \sigma^2}} \right)\sin \left(2 \pi f\left(t-t_0\right)\right)
\label{eq:field_profile}
\end{equation}
Where $H$ is the magnitude of the Gaussian envelope, $\sigma$ is the standard deviation, and $f$ is the frequency of the pulse. The exchange for Mn\textsubscript{2}Au consists of the 13 interactions for each site of which 9 are AFM coupling and 4 correspond to coupling in the FM planes. This set of exchange parameters is the same as used in Ref. \cite{hirst_mn2au} and yields a N\'eel temperature of $\approx 1350$ K. Experimental work puts the N\'eel temperature between 1300 and 1600 K \cite{Barthem2013}.
\section{Reversal in Mn\textsubscript{2}Au}
\label{sec:mn2au}
\begin{figure}
    \centering
    \includegraphics[scale=0.7]{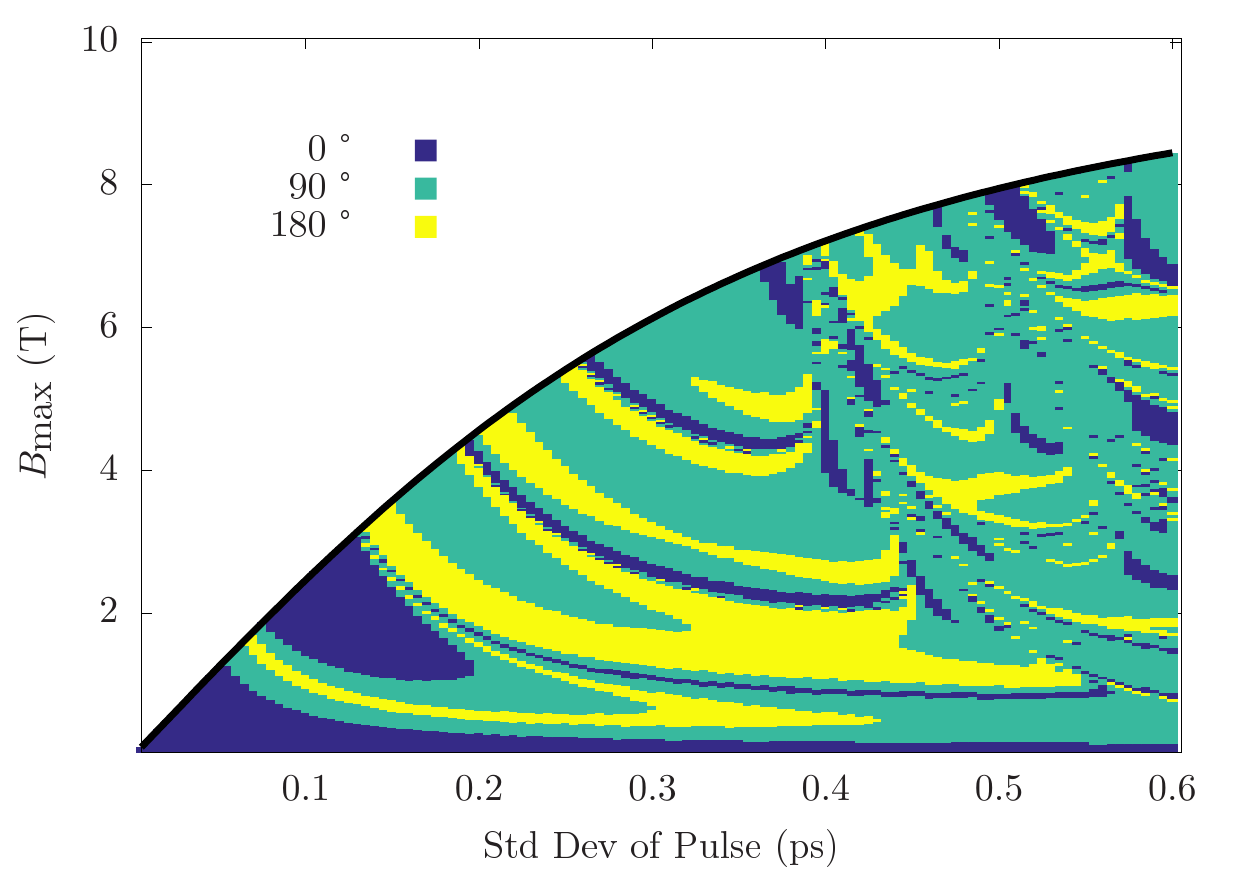}
    \caption{Switching phase diagram for pure Mn\textsubscript{2}Au following a THz pulse. Turquoise and yellow regions show regions of 90\textdegree \ and 180\textdegree \ switching respectively.}
    \label{fig:0K_mn2au_map}
\end{figure}
We firstly consider a system of pure Mn\textsubscript{2}Au subject to a staggered field at 0 and 300 K. Initial calculations (not presented here) show that the switching at T=0 K is precessional and therefore we can simulate  a single unit cell. However, in the room temperature simulations we simulate $70 \times 70 \times 70$ unit cells of Mn\textsubscript{2}Au, each consisting of 4 atoms yielding a total system size of 1,372,000 Mn atoms with periodic boundary conditions at every surface. We simulate 9 values of $H$ in Eq. \eqref{eq:field_profile} varying from 80 to 720 mT in steps of 80 mT. It is worth mentioning that $H$ does not correspond to the peak field amplitude, $B_{\textrm{max}}$. A value of $H = 720$ mT for $\sigma  = 0.3$ ps yields a maximum field amplitude of $B_{\textrm{max}} \approx 442$ mT. The standard deviation and frequency remain fixed in this instance at $\sigma = 0.3$ ps and $f = 0.67$ THz respectively. Fig. \ref{fig:angle_300K} shows the N\'eel vector reorientation in the easy-plane following the THz pulse pulse at 0 K (left pane) and 300 K (right pane). It is worth noting that the out-of-plane angle remains close to zero through the entire process due to the large uniaxial easy-plane anisotropy constant $d_z$ in Eq. \eqref{eq:hamil_Mn}. The onset of 90\textdegree \ switching occurs  $B_{\mathrm{max}} = 294 $ mT with 180\textdegree \ being observed at $B_{\textrm{max}} = 442$ mT for $T = 0$ K. For 300 K, the simulations were repeated 8 times for all field values to account for the stochastic thermal effects. Dotted lines show the results from individual simulations, and shaded area represents the upper and lower bound for each field value. Upon comparing both panes in Fig. \ref{fig:angle_300K}, the first instance of reversal has reduced from $B_{\textrm{max}} = 294$ mT for 0 K, to $B_{\textrm{max}} = 245$ mT for 300 K due to the reduction in anisotropy field. In the room temperature simulations, the switching was deterministic; however, the path and final reorientation angle were not - as seen for a field amplitude of $B_{\mathrm{max}} = 392$ mT, where both 90\textdegree \ and 180\textdegree \ switching occurred. In the study by Selzer \textit{et al.} \cite{selzer_mn2au}, much smaller switching fields of $\approx 76$ mT were required for deterministic switching close to room temperature although it is worth noting that this is as a result of a 'step-like' staggered field compared the cyclic field profiles used in Fig. \ref{fig:angle_300K}.
\begin{figure}
    \centering
    \includegraphics[scale=0.35]{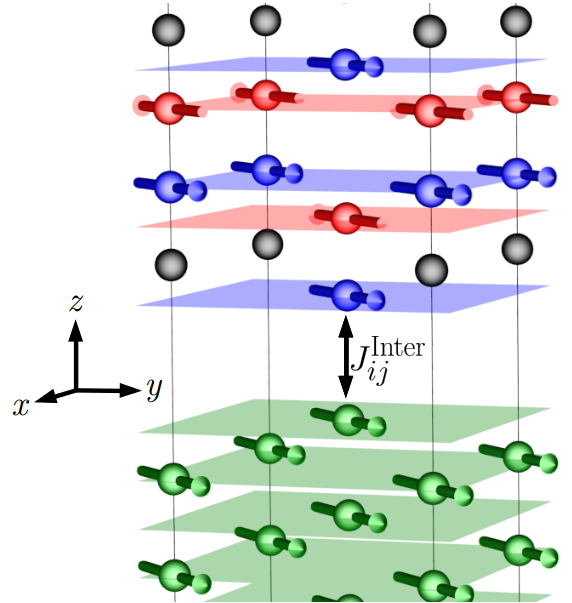}
    \caption{A snapshot of the system close to the interface. The Py atoms are shown in green, the Au in black and the Mn sublattices in blue and red respectively. The Py and Mn is coupled ferromagnetically at the interface. Each axis has been scaled differently to aid in the visualisation.}
    \label{fig:system}
\end{figure}
\begin{figure*}
    \includegraphics[width=1.0\linewidth]
    {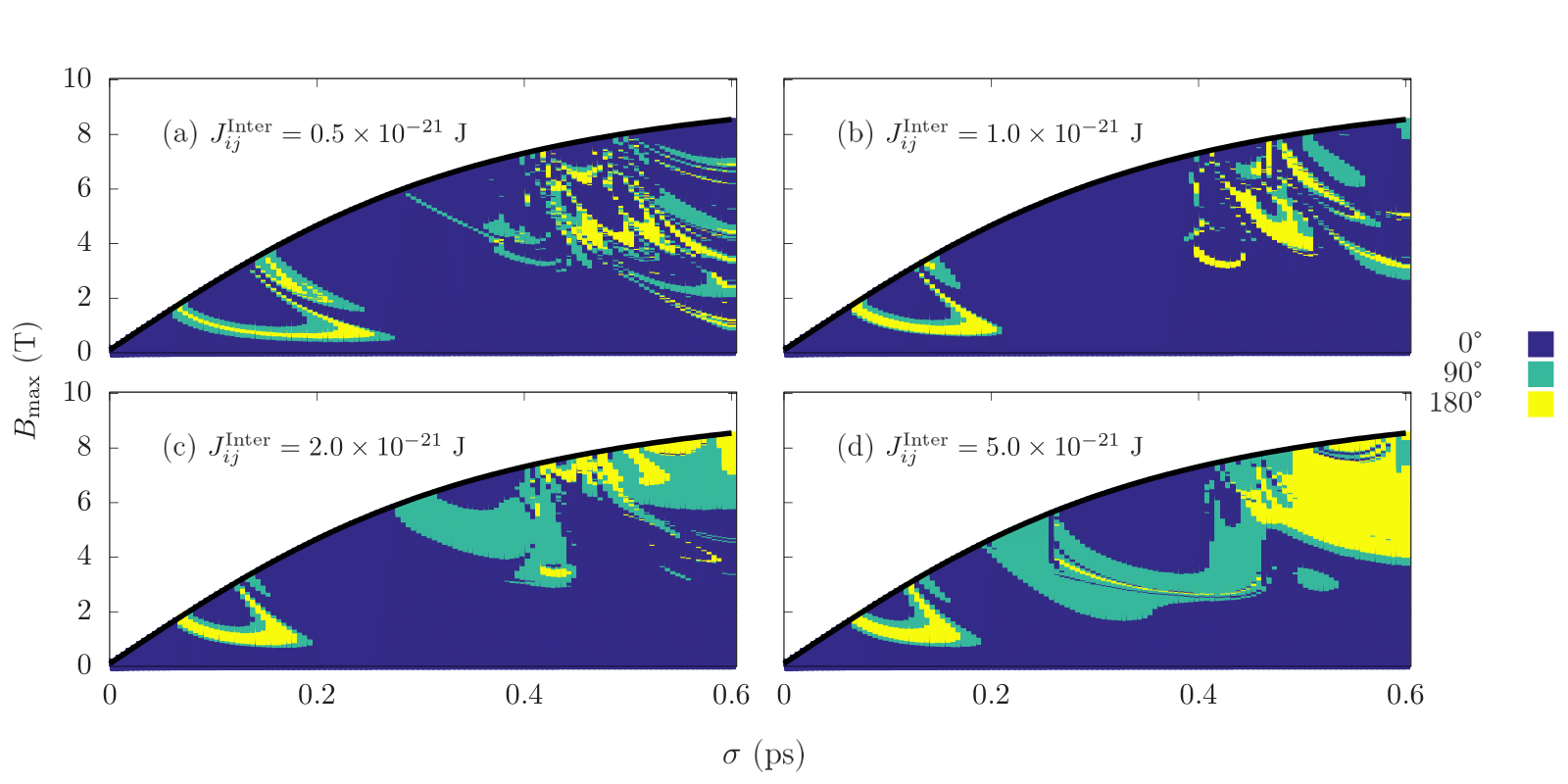}
                \subfloat{%
        \label{fig:vary_coupling:a}%
    }
       \subfloat{%
        \label{fig:vary_coupling:b}%
    }
        \subfloat{%
        \label{fig:vary_coupling:c}%
    }
       \subfloat{%
        \label{fig:vary_coupling:d}%
    }
    \caption{Magnetization reversal using a THz pulse for a permalloy thickness of 6 unit cells for different coupling strengths between FM and AFM. (a) $0.5 \times 10^{-21}$ J, (b)  $1.0 \times 10^{-21}$ J, (c)  $2.0 \times 10^{-21}$ J, and (d) $5.0 \times 10^{-21}$ J. Turqoise and yellow are regions of 90\textdegree \ and 180\textdegree \ switching respectively.}
    \label{fig:vary_coupling}
\end{figure*}
\\
\\
Previous attempts to characterise the staggered spin orbit field in terms of an electric field give varying strengths. In Ref. \cite{selzer_mn2au}, Electric fields of $10^7$ V/m yield staggered fields of about 76 mT. Much lower magnitudes of the staggered field were calculated in Refs. \cite{roy_mn2au, PhysRevApplied.9.064040} but orbital contributions were not taken into account. For the THz staggered switching fields of around $\approx 300$ mT discussed in the previous paragraph, the electric fields required to generate such a field should be achievable in experimental facilities \cite{PhysRevApplied.19.034018, 10.1063/5.0057511, 10.1063/5.0080357}.
\\
\\
For precessional switching, it is possible to describe the easy-plane magnetization dynamics of Mn\textsubscript{2}Au using a single equation of motion \cite{Gomonay2018}
\begin{equation}
\label{eq:diff_eq}
\frac{1}{\gamma H_{\mathrm{ex}}} \ddot{\varphi}_{\mathrm{L}}+2 \alpha_{\mathrm{G}} \dot{\varphi}_{\mathrm{L}}+\gamma H_{\mathrm{an}} \sin \left(4 \varphi_{\mathrm{L}}\right)=-\lambda_{\mathrm{NSOT}} \sigma_c B \cos \left(\varphi_{\mathrm{L}}\right)
\end{equation}
where $\varphi_\mathrm{L}$ is the angle of rotation in the easy plane, $B$ is the staggered field given in equation \eqref{eq:field_profile}, $\alpha_{\mathrm{G}}$ is the Gilbert damping constant, $H_{\mathrm{an}}$ and $H_{\mathrm{ex}}$ are the anisotropy and exchange fields respectively, $\sigma_c$ is the conductivity and $\lambda_{\mathrm{NSOT}}$ is the torquance from the staggered field. By fitting the 0 K magnetization dynamics shown in Fig \ref{fig:angle_300K} to Eq. \eqref{eq:diff_eq}, a value for $\lambda_{\mathrm{NSOT}}$ can be extracted and compared directly to experimental results irrespective of the field profile.  Fitting to the 0K dynamics, using a conductivity of $1.5 \times 10^{5} \mathrm{~m}^{-1} \mathrm{~V}^{-1} \mathrm{~A}$ \cite{Gomonay2018} a $\lambda_{\mathrm{NSOT}} = 234 \mathrm{~A}^{-1} \mathrm{~s}^{-1} \mathrm{~cm}^2$. A selection of the fits used to extract $\lambda_{\mathrm{NSOT}}$ are shown by the empty circles in the left hand pane of Fig. \ref{fig:angle_300K}. A more detailed discussion of the fitting process can be found in Supplementary Information S1.
\\
\\
For 0 K, it is possible to sample a large phase space of $B_{\mathrm{max}}$ and $\sigma$ with little computational cost. In Fig. \ref{fig:0K_mn2au_map} we present a switching phase diagram at $T=0$K for a range of applied field amplitudes and pulse durations for a fixed frequency of 0.67 THz. The color shows whether the system undergoes 90 or 180 degree switching (relative to the initial state) following a THz staggered pulse. The field and pulse duration required for N\'eel vector reversal increases for a multi-cycle signal because of the changing sign of the external field. For $\sigma \gtrsim 0.4$ ps, the switching window appears less structured, which is likely due to the transition from a single-cycle to a multi-cycle field with increasing $\sigma$. What is surprising is that pulses as short as 0.2 ps with sub-Tesla staggered fields can drive antiferromagnetic switching. The phase diagram is more complex than that seen previously for antiferromagnetic NiO \cite{PhysRevLett.108.247207} due to the use of an un-staggered square field profile with a single uniaxial anisotropy constant in that study. 


\section{Reversal in Mn\textsubscript{2}Au/Py Bilayers}
\label{sec:mn2au_py}
\begin{figure*}

    \includegraphics[width=1.0\linewidth]
    {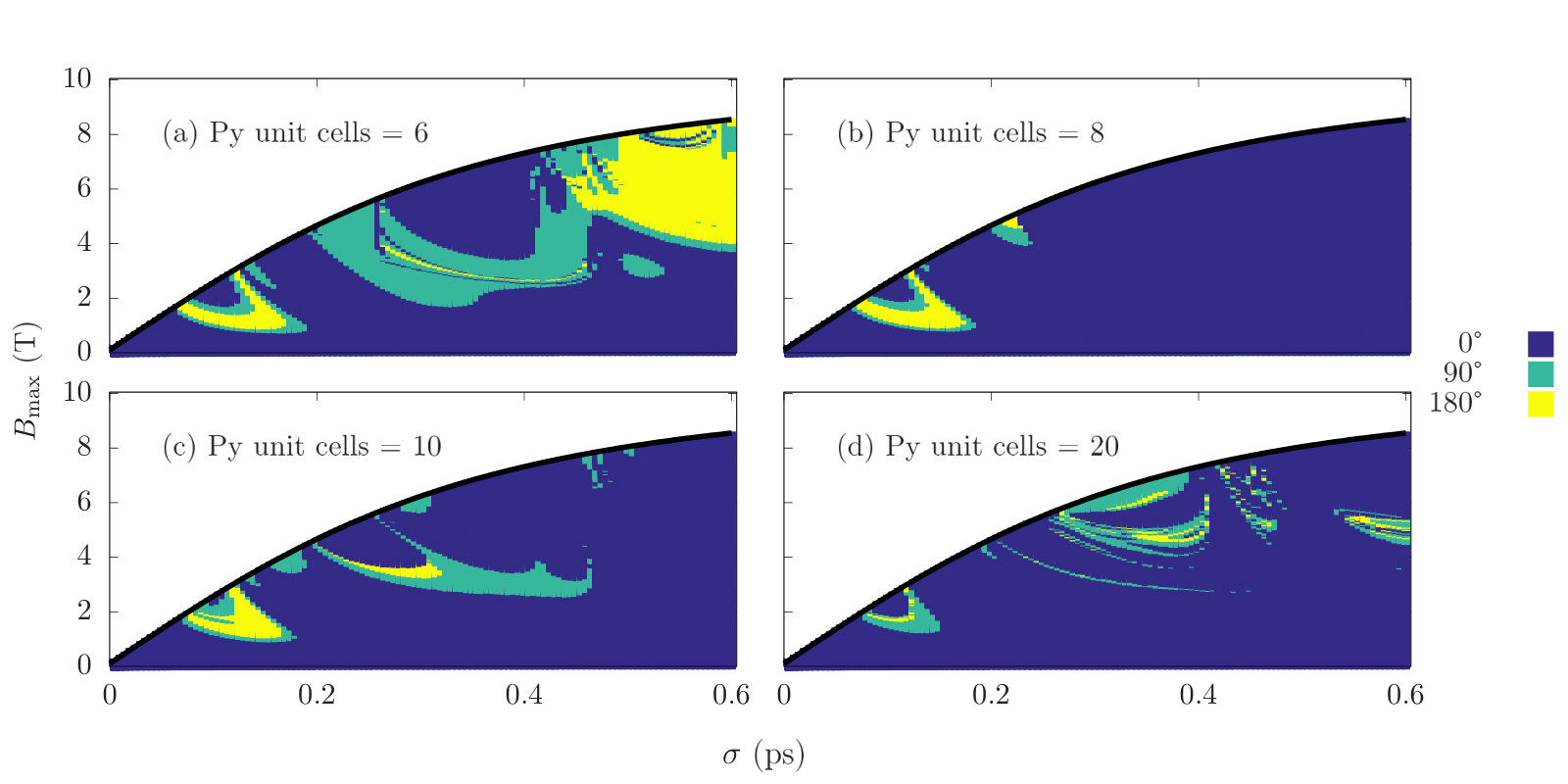}
            \subfloat{%
        \label{fig:vary_thickness:a}%
    }
       \subfloat{%
        \label{fig:vary_thickness:b}%
    }
        \subfloat{%
        \label{fig:vary_thickness:c}%
    }
       \subfloat{%
        \label{fig:vary_thickness:d}%
    }
    \caption{Magnetization reversal using a THz pulse for an interface exchange of $5.0 \times 10 ^{-21}$ J with (a) 6, (b) 8, (c) 10 and (d) 20 unit cells of Py at T = 0 K. Turquoise and yellow are regions of 90\textdegree \ and 180\textdegree \ switching respectively.}
    \label{fig:vary_thickness}
\end{figure*}
\begin{figure*}   
    \includegraphics[width=1.0\linewidth]{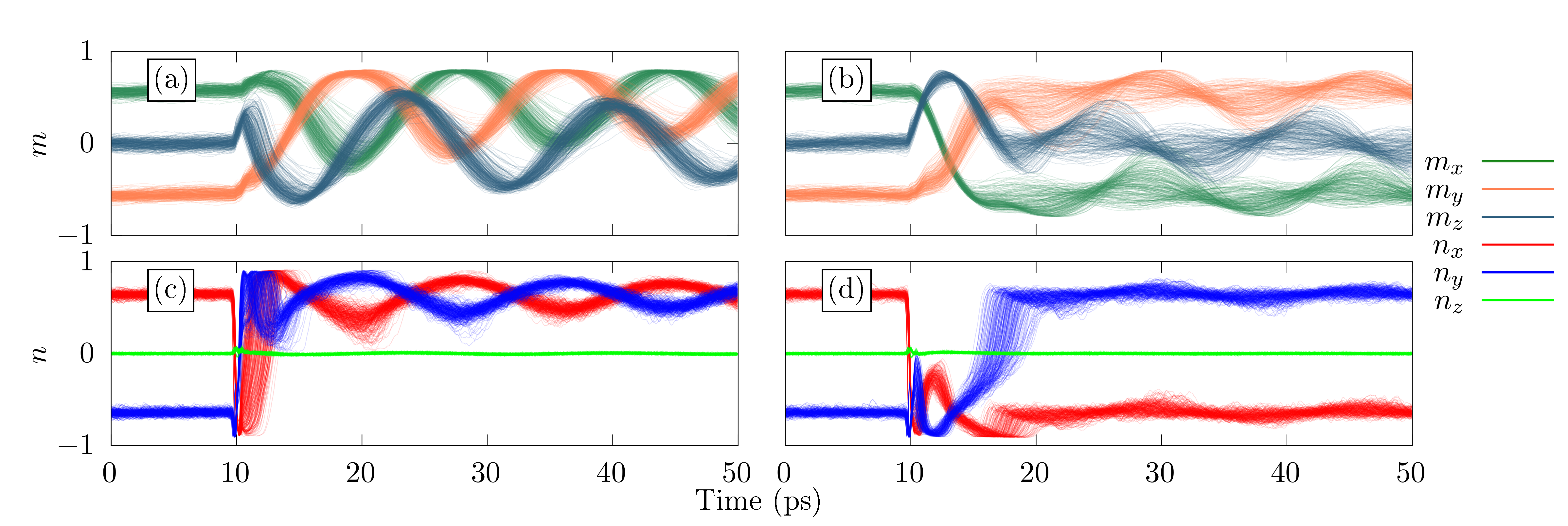}
        \subfloat{%
        \label{fig:300K_dynamics:a}
    }
       \subfloat{%
        \label{fig:300K_dynamics:b}
    }
    \caption{Magnetization dynamics at 300 K for the two most common reversal paths. (a) an anti-clockwise rotation of 270\textdegree \ in the $xy$-plane for the Mn sublattices and a clockwise rotation of 90\textdegree \ for the Py. (b) an anti-clockwise reorientation of 180\textdegree \ in the $xy$ plane for Mn sublattices and a rotation of 180\textdegree out of the $xy$ plane such that the Py sublattice is parallel to the $z$ axis roughly halfway through the switching process. These two paths account for 52\% of all switching processes at 300 K.}
    \label{fig:300K_dynamics}
\end{figure*}

Now that we have established the feasibility of ultrafast THz switching in pure Mn\textsubscript{2}Au, in this section we demonstrate magnetization reversal of ferromagnetically coupled Mn\textsubscript{2}Au/Py bilayer structures. The introduction of Py requires further material parameters, increasing the complexity of the parameter space. Here we focus on: (i) thickness of the Py (ii) thickness of Mn\textsubscript{2}Au and (iii) the coupling at the interface. In this work, we keep the thickness of Mn\textsubscript{2}Au fixed at 25 units cells along the length of the system equating to roughly 21 nanometers (nm). We chose to vary the thickness of Py using 6, 8, 10 and 20 unit cells. This corresponds to thin film thicknesses of approximately 2.13 nm, 2.84 nm, 3.55 nm and 7.1 nm respectively - similar thicknesses of Py used in the Mn\textsubscript{2}Au/Py bilayer experiments of Bommanaboyena \textit{et al.} \cite{Bommanaboyena2021}. We include open boundary conditions at either end of the chain and periodic boundaries on all others. The Py is stacked on top of the Mn2Au in the (001) direction such that the the same AFM sublattice couples to the FM at the interface. For Mn\textsubscript{2}Au we use lattice constants of $a_{\mathrm{Mn\textsubscript{2}Au}} = 3.33$ \text{\normalfont\AA} and $c_{\mathrm{Mn\textsubscript{2}Au}} = 8.537$ \text{\normalfont\AA}. We also assume the Py matches $a_{\mathrm{Mn\textsubscript{2}Au}}$ given a comparable value of $a_{\mathrm{Py}} = 3.55$ \text{\normalfont\AA} for bulk Permalloy. A schematic of the chain close to the interface can be found in Fig. \ref{fig:system}. 
\\
\\
 For the Py, we use a simplistic model with nearest-neighbour (n.n.) totalling 12 interactions per Py site. At the interface, exchange is treated as nearest neighbour at the two neighbouring atoms at the interface. In the simulations, the exchange coupling at the interface is varied from $0.5 \times 10^{-21}$ J ($\approx15\%$ of the n.n $J_{ij}$ used for Py) to $5.0 \times 10^{-21}$ J. The interface coupling is chosen as a variable quantity in our simulations as it can be manipulated experimentally by the insertion of a non-magnetic spacer between the FM and AFM \cite{Cuadrado2018} or doping at the FM/AFM interface. For the simulations of Mn\textsubscript{2}Au and Permalloy bilayers, there are two additional Hamiltonian terms:
\begin{equation}
    \mathcal{H} = \mathcal{H}_{\textrm{Mn}} + \sum_{\textrm{Py}} J_{i j}^{\textrm{Py}} \boldsymbol{S}_i \cdot \boldsymbol{S}_j + \sum_{\textrm{Inter}} J_{i j}^{\textrm{Inter}} \boldsymbol{S}_i \cdot \boldsymbol{S}_j
\end{equation}
The first term in the above can be found in Eq. \eqref{eq:hamil_Mn}. For the Permalloy, we use an average moment of $0.95 \mu_B$ with nearest neighbour exchange and exclude any anisotropy. For the 12 neighbouring interactions in Py, we use a value of $J_{i j}^{\textrm{Py}} = 3.01 \times 10^{-21}$ J which yields a Curie temperature $\approx 720$ K. 
\\
\\
The applied field is once again staggered for each Mn sublattice. An identical field to the sublattice containing the Mn atom at the interface is applied to the Py sublattice. Firstly, we consider the case of Py at 6 unit cells, and vary the strength of the coupling across the interface. Fig. \ref{fig:vary_coupling} shows four different coupling strengths, namely 0.5, 1.0, 2.0 and $5.0 \times 10^{-21}$ J. For low $\sigma$, the phase diagram varies minimally in structure across an order of magnitude increase in the interface exchange. While this region remains similar, there is a significant change in the phase diagram for $\sigma \gtrsim 0.2$ which transitions from small and scattered switching windows, to larger more continuous regions of magnetization reversal. Simulations are also conducted for constant exchange coupling with varied Py thickness, we chose the largest value of $ J_{i j}^{\textrm{Inter}} = 5.0 \times 10^{-21}$ J and vary the thickness from 6 to 20 unit cells, similar in range to the bilayer experiments in Ref. \cite{Bommanaboyena2021}. As was seen in Fig. \ref{fig:vary_coupling}, the phase diagram remains similar in shape for $\sigma \lesssim 0.2$, but there is a notable increase in the fields required for reversal in this region with almost no 180\textdegree \ reversal observed for the case of 20 Py unit cells. For $\sigma \gtrsim 0.2$, the change in the phase diagram as we transition from 6 to 20 unit cells is more disordered. In contrast to switching in pure Mn\textsubscript{2}Au, there exist large bands in Figs \ref{fig:vary_coupling} and \ref{fig:vary_thickness} where no magnetization reversal is observed, which suggests the FM is hindering switching.
\begin{figure*}
\hspace*{-0.7cm}
    \includegraphics[width=1.0\linewidth]{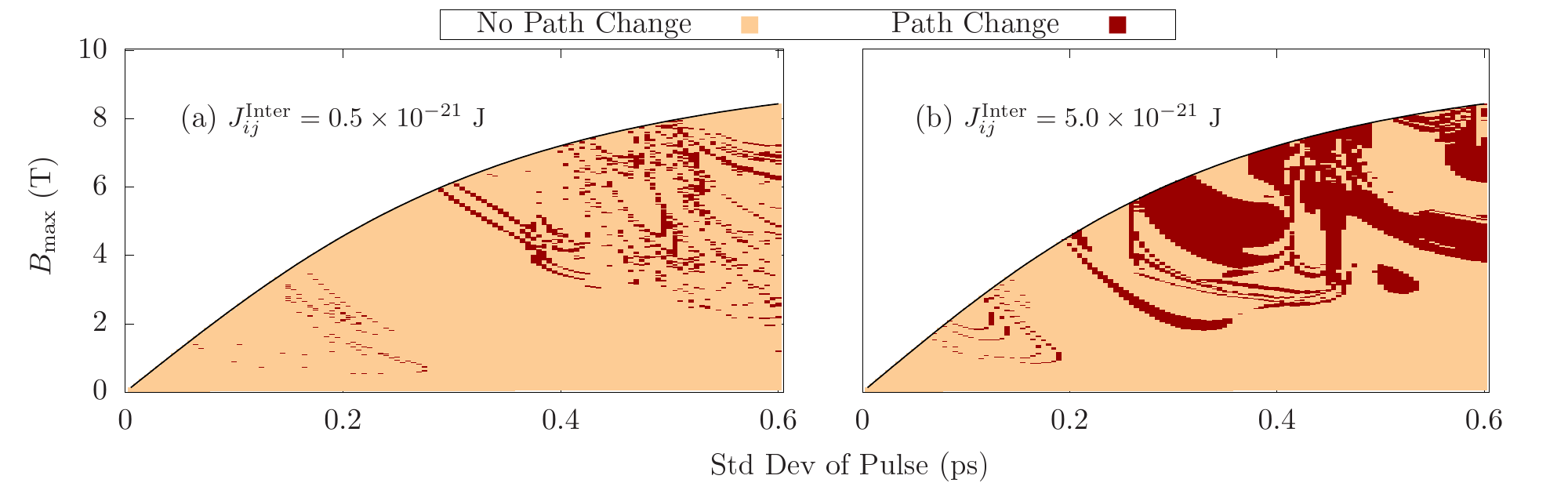}
                \subfloat{%
        \label{fig:mn2au_vs_mn2au+py:a}%
    }
       \subfloat{%
        \label{fig:mn2au_vs_mn2au+py:b}%
    }
    \caption{Changes in phase diagram from an excitation of both Mn and Py sublattices to just Mn sublattices for coupling strengths of (a) $0.5 \times 10^{-21}$ J and (b) $5.0 \times 10^{-21}$ J. Dark blue areas represent areas of no change in switching between the two cases. Other possible scenarios can be found in the colorbar.}
    \label{fig:mn2au_vs_mn2au+py}
\end{figure*}
\begin{figure*}
    \includegraphics[width=1.0\linewidth]{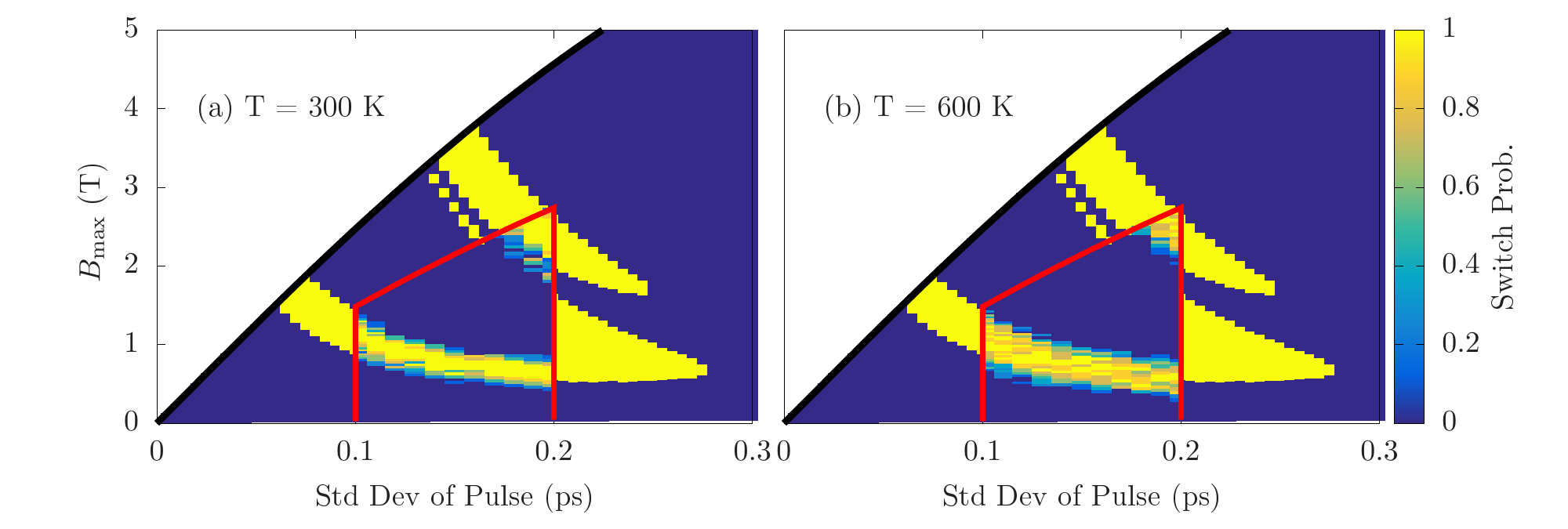}
    \caption{Magnetization reversal using a multicycle pulse for a Py thickness of 6 unit cells with a coupling strength of $0.5 \times 10^{-21}$ J. The area inside the red rectangle are simulations at finite temperature. With (a) $T = 300$ K and (b) $T = 600$ K on the left and right respectively. The switching probability is averaged over 8 repeated simulations. Area outside the box is at 0 K, as shown in Fig. \ref{fig:0K_mn2au_map}. There is no differentiating between 90\textdegree \ and 180\textdegree \ reversal.}
    \label{fig:0K_vs_300K}
\end{figure*}
\\
\\
To further reiterate this point, simulations were conducted where only the Mn sublattices were subject to an applied staggered field. Fig. \ref{fig:mn2au_vs_mn2au+py} shows the change in phase diagram as a result of an excitation of the entire system compared to the case of excitation only in the AFM. For $\sigma \lesssim 0.2$ ps, there is little observed change between Fig \ref{fig:vary_coupling:a} and Fig \ref{fig:mn2au_vs_mn2au+py:a}, as well as \ref{fig:vary_coupling:d} and Fig \ref{fig:mn2au_vs_mn2au+py:b}. Changes are only observed at the boundary between two reorientation angles by, in most cases, a  single shift in the field strength or pulse width increment. The AFM is therefore almost almost entirely responsible for the switching in this region. For larger pulse widths, the FM is given more time to respond to the applied field and therefore the role of the field in the FM becomes more important as shown by the large changes in reorientation for $\sigma \gtrsim 0.3$.
\\
\\
It's well understood that elevated temperatures lead to a reduction in anisotropy fields \cite{CALLEN19661271, PhysRevB.102.020412, Mryasov2005}. Such heating is the key principle behind Heat Assisted Magnetic Recording (HAMR) \cite{hamr1, hamr2, hamr3, hamr4}. Here, we conduct finite temperature simulations for the bilayer containing 6 unit cells of Py along the chain with an interface exchange of $0.5 \times 10^{-21}$ J - which shows the most favorourable switching characteristics. Due to the increased computational cost of finite temperature ASD simulations a smaller phase space is sampled. We simulate in the ranges $0.1 \leq \sigma \leq 0.2$ ps and $0 \leq H \leq 6$ T (recall $H$ does not correspond to the maximum amplitude, $B_\mathrm{max}$). Simulations are repeated 8 times for each field strength and pulse width. Fig. \ref{fig:0K_vs_300K} shows the switching probability for the two aforementioned temperatures where there is a lowering and broadening to the switching band transitioning from 300 to 600 K. At 300 K, the equilibrium magnetization is roughly $0.84 \ M/M_S$ and $0.92 \ M/M_S$ for the Py and Mn\textsubscript{2}Au sublattices respectively. At 600 K, it is approximately $0.42 \ M/M_S$ and $0.81 \ M/M_S$.  The area inside the red box shows the finite temperature probability. The area outside the red box shows switching at 0 Kelvin. The switching is always deterministic at 0K as there is no stochastic noise processes. There is no differentiating between 90\textdegree \ and 180\textdegree \ reversal in Fig. \ref{fig:0K_vs_300K}; what is shown is the the probability of either event occurring.
\\
\\
The reversal path for the Mn sublattices remains circular in the $xy$-plane across all temperatures because of the large negative uniaxial anisotropy constant $d_z$. The Py reversal path is also circular with no significant reduction in the magnetization following the application of the field. However, interestingly, the lack of anisotropy allows for the precession of the Py sublattice out of the $xy$-plane. 
Fig. \ref{fig:300K_dynamics} shows the magnetization dynamics for the two most common reversal paths for an interface exchange of $0.5 \times 10^{-21}$ J with a Py thickness of 6 unit cells. Because of the highly circular precession in the x-y plane resulting from the strong uniaxial anisotropy $d_z$, the reversal pathways are derived from the dynamics of the Mn sublattices, making it simple to characterise the path in terms of a rotation about the $z$ axis. Fig. \ref{fig:300K_dynamics:a} shows the dynamics corresponding to a 270\textdegree \ clockwise rotation of the Mn sublattices and \ref{fig:300K_dynamics:b} shows a reorientation after a clockwise rotation of 180\textdegree. In total, these two paths account for 51\% of all the switching events at 300 K with \ref{fig:300K_dynamics:a} and \ref{fig:300K_dynamics:b} accounting for 29\% and 22\% of the switching events, respectively. If we characterise the reversal time as the time taken for the Mn sublattices to reach the final steady state, the average reversal time is $\sim 5$ ps and $\sim 9$ ps for Fig. \ref{fig:300K_dynamics:a} and \ref{fig:300K_dynamics:b} respectively. 
There exist 8 other paths with occurrences of between 1\% and 9\% that account for the the remaining 49\% of switching events.  While the switching events occur in only a few picoseconds, the system continues to oscillate at its resonant frequency for an extended period of time. The frequency of this resonance is characterised by both the thickness of the Py layer and the strength of the interface coupling. Increasing the thickness of Py reduces the resonant frequency, whereas increasing the coupling strength, $ J_{i j}^{\textrm{Inter}}$, at the interface causes an increase in the resonant frequency. Both of these quantities could act as a way to tune to desired frequencies for experimental applications. Further discussion of the resonant frequency can be found in Supplementary Information section S2.
\section{Discussion}
Here we have presented results for magnetization reversal in pure Mn\textsubscript{2}Au as well as an FM/AFM bilayer consisting of an average moment model of Permalloy and Mn\textsubscript{2}Au. We have shown that reversal across the bilayer is possible as a result of sub-picosecond THz fields applied in an in-plane direction that maximises the torque, with amplitudes no larger than a few Tesla. The magnetization dynamics have been used to accurately parameterise the torquance, $\lambda_{\mathrm{NSOT}}$, which can be directly compared to experiment irrespective of the field profile. The importance of these findings is based on the ability to generate staggered fields in Mn\textsubscript{2}Au via NSOTs and, more practically, create FM/AFM bilayers which have one-to-one mapping of the N\'eel vector on the FM domain structure as seen in Ref. \cite{Bommanaboyena2021}.  In this work we have focused on several important physical parameters including  the field amplitude, field standard deviation, the interface exchange and the thickness of the FM layer. It is also worth noting that the FM used in the bilayer system should be treated as generic and could easily be interchanged with another weakly anisotropic material. Nonetheless, we have shown that reversal in FM/AFM Bilayer systems are possible using sub-picosecond fields with amplitudes of $\sim 2$ Tesla. Regions of switching have been identified where staggered fields in the AFM alone are responsible for magnetization reversal. Simulations at finite temperature show a small reduction in the fields required for switching. A choice of FM with a similar critical temperature to Mn\textsubscript{2}Au would allow switching at further elevated temperatures, likely resulting in further reductions of the required field amplitudes.
\section{Acknowledgements}
This work was supported by the EPSRC TERASWITCH project  (Project  ID  EP/T027916/1). Simulations were completed using resources provided by the Cambridge Tier-2 system operated by the University of Cambridge Research Computing Service (www.hpc.cam.ac.uk) funded by EPSRC Tier-2 capital grant EP/T022159/1 as well as the Baskerville Tier 2 HPC service. Baskerville was funded by the EPSRC and UKRI through the World Class Labs scheme (EP/T022221/1) and the Digital Research Infrastructure programme (EP/W032244/1) and is operated by Advanced Research Computing at the University of Birmingham.

\newpage
\bibliography{ref}
\end{document}